%Paper: hep-ph/9311312
%From: FARAGGI@sns.ias.edu
%Date: 18 Nov 1993 13:43:30 -0400 (EDT)

%
% Printing instructions:
%       This paper needs the macro packages phyzzx.tex and tables.tex
%       table 1. in \special{landscape} mode
%       should be stripped off and printed separately.
%
\input phyzzx
\tolerance=1000
\sequentialequations
\def\rl{\rightline}

\def\t1{{\tilde 1}}

\def\AEF{A.E. Faraggi}
\def\DVN{D. V. Nanopoulos}

\def\NPB#1#2#3{Nucl. Phys. B{\bf#1} (19#2) #3}
\def\PLB#1#2#3{Phys. Lett. B{\bf#1} (19#2) #3}
\def\PRD#1#2#3{Phys. Rev. D{\bf#1} (19#2) #3}
\def\PRL#1#2#3{Phys.  Rev.Lett.{\bf#1} (19#2) #3}

\REF\heterotic{D.J. Gross, J.A. Harvey, E. Martinec and
R. Rohm, \PRL{54}{85}{502}; \NPB{256}{85}{253}.}
\REF\CHSW{P.Candelas, G.T. Horowitz, A. strominger and E. Witten,
\NPB{258}{85}{46}.}
\REF\DHVW{L. Dixon, J.A. Harvey, C. Vafa and
E. Witten, \NPB{261}{85}{678}, \NPB{274}{86}{285}.}
\REF\N{K.S. Narain, \PLB{169}{86}{41}.}
\REF\NSW{K.S. Narain, Sarmadi and E. Witten, \NPB{279}{87}{369}.}
\REF\FFF{I. Antoniadis, C. Bachas, and C. Kounnas,
\NPB{289}{87}{87}; I. Antoniadis and C. Bachas,
\NPB{298}{88}{586}; H. Kawai, D.C. Lewellen, and S.H.-H. Tye,
\PRL{57}{86}{1832}; \PRD{34}{86}{3794};
\NPB{288}{87}{1}; R. Bluhm, L. Dolan, and P. Goddard,
\NPB{309}{88}{33}.}
\REF\REVAMP{I. Antoniadis, J. Ellis, J. Hagelin, and \DVN, \PLB{231}{89}{65}.}
\REF\FNY{\AEF, D.V. Nanopoulos and K. Yuan, \NPB{335}{90}{347}.}
\REF\ALR{I. Antoniadis, G. K. Leontaris and J. Rizos, \PLB{245}{90}{161}.}
\REF\EU{\AEF, \PLB{278}{92}{131}; \PLB{274}{92}{47}.}
\REF\SLM{\AEF, \PRD{47}{93}{5021}; \AEF, \NPB{387}{92}{239},
hep-th\\9208024.}
\REF\CKM{\AEF, E. Halyo, \PLB{307}{93}{305}, hep-ph\\9301261;
WIS-93-35-PH, to appear in Nucl. Phys. {\bf B},  hep-ph\\9306235.}
\REF\naturalness{A.E. Faraggi and D.V. Nanopoulos, \PRD{48}{93}{3288}.}
\REF\NRT{\AEF, \NPB{403}{93}{101}, hep-th\\9208023.}
\REF\FM{\AEF, \NPB{407}{93}{57, hep-ph\\9301220}.}
\REF\NSV{K.S. Narain, Sarmadi and C. Vafa, \NPB{288}{87}{551}.}

\singlespace
\rl{IASSNS--HEP--93/77}
\rl{\today}
\rl{T}
\normalspace
\smallskip
\titlestyle{\bf{$Z_2\times Z_2$ orbifold compactification as the origin
of realistic free fermionic models}}
\author{Alon E. Faraggi
{\footnote*{e--mail address: faraggi@sns.ias.edu}}}
\smallskip
\centerline {School of Natural Sciences}
\centerline {Institute for Advanced Study}
\centerline {Olden Lane}
\centerline {Princeton, NJ 08540}
\bigskip
\titlestyle{ABSTRACT}

I discuss the correspondence between realistic four dimensional
free fermionic models and
$Z_2\times Z_2$ orbifold compactification.
%I illustrate this correspondence for one particular example.
I discuss the properties of the $Z_2\times
Z_2$ orbifold that are reflected in the realistic free fermionic models.
I argue that the properties of the realistic free fermionic models
arise due to the underlying $Z_2\times Z_2$ orbifold compactification
with nontrivial background fields.
I suggest that three generation is a natural outcome of $Z_2\times Z_2$
orbifold compactification with ``standard embedding'' and at the point
in compactification space that corresponds to the free fermionic
formulation. I discuss how quark flavor mixing is related to the
compactification.

\nopagenumbers
\pageno=0
\singlespace
\vskip 0.5cm
\endpage
\normalspace
\pagenumbers

\centerline{\bf 1. Introduction}

Superstring theory may consistently unify gravity with the
gauge interactions. The consistency of superstring theory imposes a
certain number of degrees of freedom.
In the closed heterotic string [\heterotic], of the 26 right--moving bosonic
degrees of freedom, 16 are compactified on a flat torus and produce the
observable and hidden gauge groups. Six right--moving bosonic degrees of
freedom, combined with six left--moving degrees of freedom, are compactified
on Calabi--Yau manifold [\CHSW] or on an orbifold [\DHVW].
Alternatively, all the extra degrees of freedom, beyond the four space--time
dimensions, can be taken as bosonic [\N,\NSW], or fermionic [\FFF],
internal degrees of freedom propagating on the string world--sheet.
The different interpretations are expected to be related.
In this paper, I discuss the correspondence between realistic models in the
free fermionic formulation and $Z_2\times Z_2$ orbifold compactification.

Models constructed in the free fermionic formulation produced
the most realistic superstring models to date [\REVAMP--\FM].
In Ref. [\naturalness] it was
shown that the reduction of the number of chiral generations to three
generations is correlated in these models with the factorization of the gauge
group to observable and hidden sectors and with the breaking
of nonabelian horizontal gauge groups to factors of $U(1)$'s at most. It was
consequently argued that three generation is the most natural number
of generations in this class of models. In Refs. [\EU-\CKM]
it was suggested that the generation mass hierarchy and the suppression
of mixing terms among these generations is explained in terms of horizontal
symmetries that are derived in these models.
While the fermionic formulation enables the
construction of rather realistic models, the orbifold formulation
may relate the realistic models to the geometry at the unification scale.
An apparent example is the number of generations which is related to
the Euler characteristic in the orbifold formulation.
Thus, the two formulations have complementary merits. Therefore,
it is important to understand the connection between the two formulations.

The paper is organized as follows: in section 2, I present an
$E_6\times U(1)^2 \times E_8\times S0(4)^3$ model in the
free fermionic formulation
and its matter content. In Section 3 I show how the same model is
obtained in the orbifold language. In section 4 I discuss some of the
properties of the realistic free fermionic models and their relation to the
$Z_2\times Z_2$ orbifold.

\bigskip
\centerline{\bf 2. The model in the free fermionic formulation}

In the free fermionic formulation of the heterotic string
in four dimensions all the world--sheet
degrees of freedom  required to cancel
the conformal anomaly are represented  in terms of free fermions
propagating on the string world--sheet.
 The world--sheet
supercurrent is realized nonlinearly among the internal
left--moving free fermions,
$$T_F=\psi^\mu\partial{X_\mu}+i\chi^I{y^I}{\omega^I},{\hskip .2cm}
(I=1,\cdots,6),$$
where $X^\mu,\psi^\mu$ are the usual space--time fields and
indices, $\{\chi^I,y^I,\omega^I\}$ $(i=1,\cdots,6)$
are 18 real free fermions transforming as the adjoint representation of
$SU(2)^6$. The right--moving sector consist of ${\bar X}^\mu$
and 44 real internal free fermion fields.

Under parallel transport around a noncontractible loop of the torus
the fermionic states pick up a phase. The phases for all world--sheet fermions
are specified in 64 dimensional boundary condition vectors for all
world--sheet fermions. A model in this construction is specified
by a set of boundary condition basis vectors that spans a finite
additive group $\Xi$.
These basis vectors are constrained by the string
consistency requirements (e.g. modular invariance) and
completely determine the vacuum structure of
the model. The physical spectrum is obtained by applying the generalized
GSO projections. The low energy effective field theory is obtained
by S--matrix elements between external states. The Yukawa couplings
and higher order nonrenormalizable terms in the superpotential
are obtained by calculating correlators between vertex operators.
For a correlator to be nonvanishing all the symmetries of the model must
be conserved. Thus, the boundary condition vectors determine the
phenomenology of the models.

The six basis vectors (including the vector {\bf 1}) that generate the
model in the free fermionic formulation are
$$\eqalignno{S&=(
{\underbrace{1,\cdots,1}_{{\psi^\mu},{\chi^{1,...,6}}}},0,\cdots,0
\vert 0,\cdots,0)&(1a)\cr
\xi_1&=(0,\cdots,0\vert
{\underbrace{1,\cdots,1}_{{\bar\psi}^{1,\cdots,5},
{\eta^{1,2,3}}}},0,\cdots,0)&(1b)\cr
\xi_2&=(0,\cdots,0\vert0,\cdots,0,
{\underbrace{1,\cdots,1}_{{\bar\phi}^{1,\cdots,8}}})&(1c)\cr
b_1&=({\underbrace{1,\cdots\cdots\cdots,1}_
{{\psi^\mu},{\chi^{12}},y^{3,...,6},{\bar y}^{3,...,6}}},0,\cdots,0\vert
{\underbrace{1,\cdots,1}_{{\bar\psi}^{1,...,5},
{\bar\eta}^1}},0,\cdots,0)&(1d)\cr
b_2&=({\underbrace{1,\cdots\cdots\cdots\cdots\cdots,1}_
{{\psi^\mu},{\chi^{34}},{y^{1,2}},
{\omega^{5,6}},{{\bar y}^{1,2}}{{\bar\omega}^{5,6}}}}
,0,\cdots,0\vert
{\underbrace{1,\cdots,1}_{{{\bar\psi}^{1,...,5}},{\bar\eta}^2}}
,0,\cdots,0)&(1e)\cr}$$
with the choice of generalized GSO projections
$$c\left(\matrix{b_i\cr
                                    b_j\cr}\right)=
c\left(\matrix{b_i\cr
                                    S\cr}\right)=
c\left(\matrix{b_i\cr
                                    \xi_1,\xi_2\cr}\right)=
c\left(\matrix{\xi_i\cr
                                    \xi_i\cr}\right)=
-c\left(\matrix{\xi_1\cr
                                    \xi_2\cr}\right)=
-c\left(\matrix{1\cr
                                    1\cr}\right)=-1,\eqno(2)$$
and the others given by modular invariance.
The notation of refs. [\SLM] is used. The first four vectors
in the basis $\{{\bf 1},S,\xi_1,\xi_2\}$ generate a model with $N=4$
space--time supersymmetry with an $E_8\times E_8\times SO(12)$ gauge
group. The sector {\bf S} generates  $N=4$ space--time supersymmetry.
The first and second $E_8$ are obtained from the world--sheet
fermionic states $\{{\bar\psi}^{1,\cdots,5},{\bar\eta}^{1,2,3}\}$ and
$\{{\bar\phi}^{1,\cdots,8}\}$, respectively while $SO(12)$ is obtained
from $\{{\bar y},{\bar\omega}\}^{1,\cdots,6}$.
The Neveu--Schwarz sector
produces the adjoint representations of $SO(12)\times SO(16)\times SO(16)$.
The sectors $\xi_1$ and
$\xi_2$ produce the spinorial representation of $SO(16)$ of the
observable and hidden sectors respectively,
and complete the observable and hidden gauge groups to $E_8\times E_8$.

The vectors $b_1$ and $b_2$ break the gauge symmetry to
$E_6\times U(1)^2\times E_8\times SO(4)^3$ and $N=4$ to $N=1$
space--time supersymmetry.
Restricting $b_j\cdot S=0$mod$2$, and $c\left(\matrix{S\cr
                                                b_j\cr}\right)=\delta_{b_j}$,
for all basis vector $b_j\epsilon{B}$ guarantees the existence of
$N=1$ space--time supersymmetry. The superpartners from a given sector
$\alpha\epsilon\Xi$ are obtained from the sector $S+\alpha$.
We denote the $U(1)$ generators,
that are generated by the world--sheet currents
$:{\bar\eta}^i{\bar\eta}^{i*}:$, by $U(1)_i$.
The fermionic states $\{\chi^{12},\chi^{34},\chi^{56}\}$ and
$\{{\bar\eta}^1,{\bar\eta}^2,{\bar\eta}^3\}$ give the usual ``standard--
embedding", with
$b(\chi^{12},\chi^{34},\chi^{56})=b({\bar\eta}^1,{\bar\eta}^2,{\bar\eta}^3)$.
The $U(1)$ current of the left--moving $N=2$ world--sheet supersymmetry
is given by $J(z)=i\partial_z(\chi^{12}+\chi^{34}+\chi^{56})$.
The $U(1)$ charges in the decomposition of $E_6$ under
$SO(10)\times U(1)$ are given by $U(1)_{E_6}=U(1)_1+U(1)_2+U(1)_3$
while the charges of the two orthogonal combinations are given by
$U(1)^\prime=U(1)_1-U(1)_2$ and
$U(1)^{\prime\prime}=U(1)_1+U(1)_2-2U(1)_3$.
The three $SO(4)$ gauge groups are produced by the
right--moving world--sheet
fermionic states $\{{\bar y}^{3,\cdots,6}\}$, $\{{\bar y}^1,{\bar y}^2
{\bar\omega}^5,{\bar\omega}^6\}$ and $\{{\bar\omega}^{1,\cdots,4}\}$.

The massless spectrum of the model consist of the following sectors.
The Neveu--Schwarz and $\xi_1$ sectors produce in addition to the
spin 2 and spin 1 states three copies of chiral multiplets that
transform as $27+{\bar{27}}$ under $E_6$, and an equal number of
$E_6$ singlets that are charged under $U(1)^2$. The Neveu--Schwarz
sector also produces three scalar multiplets that transform as
$(4,{\bar 4},1)$, one under each of the horizontal $SO(4)$ symmetries.
The sectors $b_1$, $b_2$ and $b_3=1+\xi_2+b_1,b_2$ plus $b_j+\xi_2$
produce 24 chiral 27 of $E_6$, and 24 $E_6$ singlets that are
charged under $U(1)_1$, $U(1)_2$. In the decomposition of $E_6$ under
$SO(10)$, the sectors $b_j$ produce the 16 representation of $SO(10)$
while the sectors $b_j+\xi_2$ produce the $10+1$ in the 27 representation
of $E_6$.
The  sectors $b_j+\xi_1$ produce an equal number of $E_6$ singlets.
The singlet of $SO(10)$ in the 27 of $E_6$ and the additional $E_6$ singlet
from the sectors $b_j+\xi_1$ are produced by acting on the degenerate vacuum
with ${\bar\eta}_j$ and ${\bar\eta}_j^*$.
In addition to these states the sectors
$b_j+\xi_1$ produce $E_6\times E_8$ singlets which carry $U(1)^2$ charges
and that transform nontrivialy under the horizontal $SO(4)$ symmetries.

In this model the only internal fermionic states which count the
multiplets of $E_6$ are the real internal fermions $\{y,w\vert{\bar y},
{\bar\omega}\}$. This is observed by writing the degenerate vacuum
of the sectors $b_j$ in a combinatorial notation. The vacuum of the sectors
$b_j$  contains twelve periodic fermions. Each periodic fermion
gives rise to a two dimensional degenerate vacuum $\vert{+}\rangle$ and
$\vert{-}\rangle$ with fermion numbers $0$ and $-1$, respectively.
The GSO operator, is a generalized parity, operator which
selects states with definite parity. After applying the
GSO projections, we can write the degenerate vacuum of the sector
$b_1$ in combinatorial form
$$\eqalignno{\left[\left(\matrix{4\cr
                                    0\cr}\right)+
\left(\matrix{4\cr
                                    2\cr}\right)+
\left(\matrix{4\cr
                                    4\cr}\right)\right]
\left\{\left(\matrix{2\cr
                                    0\cr}\right)\right.
&\left[\left(\matrix{5\cr
                                    0\cr}\right)+
\left(\matrix{5\cr
                                    2\cr}\right)+
\left(\matrix{5\cr
                                    4\cr}\right)\right]
\left(\matrix{1\cr
                                    0\cr}\right)\cr
+\left(\matrix{2\cr
                                    2\cr}\right)
&\left[\left(\matrix{5\cr
                                    1\cr}\right)+
\left(\matrix{5\cr
                                    3\cr}\right)+
\left(\matrix{5\cr
                                    5\cr}\right)\right]\left.
\left(\matrix{1\cr
                                    1\cr}\right)\right\}&(3)\cr}$$
where
$4=\{y^3y^4,y^5y^6,{\bar y}^3{\bar y}^4,
{\bar y}^5{\bar y}^6\}$, $2=\{\psi^\mu,\chi^{12}\}$,
$5=\{{\bar\psi}^{1,\cdots,5}\}$ and $1=\{{\bar\eta}^1\}$.
The combinatorial factor counts the number of $\vert{-}\rangle$ in the
degenerate vacuum of a given state.
The two terms in the curly brackets correspond to the two
components of a Weyl spinor.  The $10+1$ in the $27$ of $E_6$ are
obtained from the sector $b_j+X$.
{}From Eq. (4) it is observed that the states
which count the multiplicities of $E_6$ are the internal
fermionic states $\{y^{3,\cdots,6}\vert{\bar y}^{3,\cdots,6}\}$.
A similar result is
obtained for the sectors $b_2$ and $b_3$ with $\{y^{1,2},\omega^{5,6}
\vert{\bar y}^{1,2},{\bar\omega}^{5,6}\}$
and $\{\omega^{1,\cdots,4}\vert{\bar\omega}^{1,\cdots,4}\}$
respectively, which suggests that
these twelve states correspond to a six dimensional
compactified orbifold with Euler characteristic equal to 48.

\bigskip
\centerline{\bf 3. The model in the orbifold formulation}

I now describe how to construct the same model in the orbifold
formulation.
In the orbifold formulation [\DHVW] one starts with a model compactified
on a flat
torus with nontrivial background fields [\NSW].
The action for the six compactified
dimensions is given by,
$$S={1\over{8\pi}}
\int{d^2\sigma({G_{ij}\partial{X^i}\partial{X_j}+B_{ij}\partial{X_i}
\partial{X_j}})}\eqno(4)$$
where
$$G_{ij}={1\over2}{\sum_{I=1}^D}R_ie_i^IR_je_j^I\eqno(5)$$
is the metric of the six dimensional compactified space
and $B_{ij}=-B_{ji}$ is the antisymmetric tensor field.
The $e^i=\{e_i^I\}$ are six linear independent vectors normalized
to $(e_i)^2=2$. The left-- and right--moving momenta are given by
$$P^I_{R,L}=[m_i-{1\over2}(B_{ij}{\pm}G_{ij})n_j]{e_i^I}^*\eqno(6)$$
where the ${e_i^I}^*$ are dual to the $e_i$, and
$e_i^*\cdot e_j=\delta_{ij}$. The left-- and right--moving momenta span a
Lorentzian even self--dual lattice. The mass formula for the left and
right movers is
$$M_L^2=-c+{{P_L\cdot{P_L}}\over2}+N_L=-1+{{P_R\cdot{P_R}}\over2}+
N_R=M_R^2\eqno(7)$$
where $N_{L,R}$ are the sum on the left--moving and right--moving oscillators
and $c$ is a normal ordering constant equal to ${1\over2}$ and $0$
for the antiperiodic (NS) and periodic (R) sectors of the NSR fermions.

For specific values of $R_I$ and for specific choices of the background
fields the $U(1)^6$ of the compactified torus is enlarged. To reproduce the
$SO(12)\times E_8\times E_8$ model of the previous section, the radius
of the six compactified dimensions is taken at $R_I=\sqrt{2}$.
The basis vectors $e_i^I$ are the simple roots of $SO(12)$. The metric
$G_{ij}$ is the Cartan matrix of $SO(12)$ and the antisymmetric tensor
field is given by,
$$B_{ij}=\cases{
G_{ij}&;\ $i>j$,\cr
0&;\ $i=j$,\cr
-G_{ij}&;\ $i<j$.\cr}\eqno(8)$$
The right--moving momenta produce the root vectors of $SO(12)$.
For $R_I=\sqrt2$ and with the chosen background fields
all the root vectors are massless, thus reproducing the same gauge group
as in the free fermionic formulation.

The orbifold model is
obtained by moding out the six dimensional torus by a discrete symmetry
group, P. The allowed discrete symmetry groups are constrained by
modular invariance. The Hilbert space is obtained by acting on the
vacuum with twisted and untwisted oscillators and by
projecting on states that are invariant under the space and group twists.
A general left--right symmetric twist is given by
$(\theta^i_j,v^i;\Theta^I_J,V^I)$ $(i=1,\cdots,6)$ $(I=1,\cdots,16)$
and $X^i(2\pi)=\theta^i_j X^j(0)+v^i$; $X^I(2\pi)=\Theta^I_J X^J(0)+V^I$.
The massless spectrum contains mass states from the untwisted and twisted
sectors. The untwisted sector is obtained by projecting
on states that are invariant under the space and group twists.
The twisted string centers around the points that are left
fixed by the space twist. In the case of ``standard embedding"
one acts on the gauge degrees of freedom in an $SU(3)\in{E_8\times E_8}$
with the same action as on the six compactified dimensions + NSR fermions.
In this case the number of chiral families (27's of $E_6$) is
given by one half the Euler characteristic,
$$\chi={1\over{\vert{P}\vert}}\sum_{g,h\in P}\chi(g,h),$$
where $\chi(g,h)$ is the number of points left fixed
simultaneously by $h$ and $g$. The mass formula for the right--movers in the
twisted sectors is given by,
$$M_R^2=-1+{{(P+V)^2}\over2}+\Delta c_\theta+N_R$$
where $V^I$ are the shifts on the gauge sector and
$\Delta c_\theta={1\over4}\sum_{k}\eta_k(1-\eta_k)$ is the contribution of
the twisted bosonic oscillators to the zero point energy and $\eta_k={1\over2}$
for a $Z_2$ twist.

To translate the fermionic boundary conditions to twists and shifts in the
bosonic formulation we bosonize the real fermionic degrees of freedom,
$\{y,\omega\vert{\bar y},{\bar\omega}\}$. Defining,
${\xi_i}={\sqrt{1\over2}}(y_i+i\omega_i)=-ie^{iX_i}$,
$\eta_i={\sqrt{1\over2}}(y_i-i\omega_i)=-ie^{-iX_i}$
with similar definitions for the right movers $\{{\bar y},{\bar\omega}\}$
and $X^I(z,{\bar z})=X^I_L(z)+X^I_R({\bar z})$.
With these definitions the world--sheet supercurrents in the bosonic
and fermionic formulations are equivalent,
$$T_F^{int}=\sum_{i}\chi_i{y_i}\omega_i=i\sum_{i}\chi_i{\xi_i}\eta_i=
\sum_{i}\chi_i\partial{X_i}.$$ The momenta $P^I$ of the compactified scalars
in the bosonic formulation are identical with the $U(1)$ charges
$Q(f)$ of the unbroken Cartan generators of the four dimensional
gauge group,
$$Q(f)={1\over2}\alpha(f)+F(f)$$
where $\alpha(f)$ are the boundary conditions of complex fermions $f$,
reduced to the interval $(-1,1]$ and $F(f)$ is a fermion number operator.

The boundary condition vectors $b_1$ and $b_2$ now translate into
$Z_2\times Z_2$ twists on the bosons $X_i$ and fermions $\chi_i$ and
to shifts on the gauge degrees of freedom. The massless spectrum
of the resulting orbifold model consist of the untwisted sector and
three twisted sectors, $\theta$, $\theta^\prime$ and $\theta\theta^\prime$.
{}From the untwisted sector we obtain the generators of the
$SO(4)^3\times E_6\times U(1)^2\times E_8$ gauge groups.
The only roots of $SO(12)$ that are invariant under the $Z_2\times Z_2$
twist are those of the subgroup $SO(4)^3$. Thus, the $SO(12)$ symmetry
is broken to $SO(4)^3$. Similarly, the shift in the gauge sector
breaks one $E_8$ symmetry to $E_6\times U(1)^2$. In addition to the gauge group
generators the untwisted sector produces: three copies of $27+{\bar{27}}$,
one pair for each of the complexified NSR left--moving fermions;
three copies of, $1+{\bar1}$, $E_6$ singlets which are charged under
$U(1)^2$. The $E_8\times E_8$ singlets are obtained from the root lattice
of $SO(12)$ and transform as $(1,4,4)$ under the $S0(4)^3$ symmetries,
one for each of the complexified NSR left--moving fermions.

The number of fixed points in each twist is 32. The total number of
fixed points is 48. The number of chiral 27's is 24, eight from each
twisted sector, and matches the number of chiral 27's in the fermionic
model. For every fixed point we obtain the $SO(4)^3\times E_6\times E_8$
singlets. These are obtained for appropriate choices of the momentum
vectors, $P^I$. The $E_6\times E_8$ singlets can be obtained by acting on the
vacuum with twisted oscillators and from combinations of the dual of the
invariant lattice, $I^*$, [\NSV].
The spectrum of the orbifold model and its symmetries are
seen to coincide with the spectrum and symmetries of the fermionic model.

\vfill
\eject
\bigskip
\centerline{\bf 4. The realistic free fermionic models}

The previous results suggest that there is a correspondence between the
models in the fermionic and orbifold formulations.
The important point to realize is that in the fermionic formulation
the 12 internal fermionic states,
$\{y,w\vert{\bar y},{\bar\omega}\}$,
correspond to the six dimensional ``compactified space''
of the orbifold. The 16 complex fermionic states
$\{{\bar\psi}^{1,\cdots,5},{\bar\eta}^{1,2,3},{\bar\phi}^{1,\cdots,8}\}$
correspond to the gauge sector, and $\chi^{1,\cdots,6}$
correspond to the RNS fermions, of the orbifold model.
The boundary conditions, assigned to the internal fermions
$\{y,w\vert{\bar y},{\bar\omega}\}$,
determine many of the properties of the low energy spectrum.
The number of generations, the presence of Higgs doublets and the
projection of Higgs triplets, the allowed cubic and quartic
order terms in the superpotential are shown to be determined
by the specific assignment of boundary conditions to these set of
internal fermions. Thus, in the realistic free fermionic models,
we learn how the internal space determines
the low energy properties of the standard model, without
an exact knowledge of what is the action (of the additional ``Wilson
line'') on the internal orbifold.

In the realistic free fermionic models the boundary condition vector $\xi_1$
is replaced by the vector $2\gamma$ in which $\{{\bar\psi}^{1,\cdots,5},
{\bar\eta}^1,{\bar\eta}^2,{\bar\eta}^3,{\bar\phi}^{1,\cdots,4}\}$
are periodic and the remaining left-- and right--moving fermionic states
are antiperiodic. The set $\{1,S,2\gamma,\xi_2\}$ generates a model
with $N=4$ space--time supersymmetry and $SO(12)\times SO(16)\times SO(16)$
gauge group. The $b_1$ and $b_2$ twist are applied to reduce the number
of supersymmetries from $N=4$ to $N=1$ space--time supersymmetry.
The gauge group is broken to
$SO(4)^3\times U(1)^3\times SO(10)\times E_8$. The $U(1)$ combination
$U(1)=U(1)_1+U(1)_2+U(1)_3$ has a non--vanishing trace and the trace of the
two orthogonal combinations vanishes. The number of generations is still
24 with a combinatorial factor for each sector $b_1$, $b_2$ and $b_3$ as in
Eq. (3). The chiral generations are now $16$ of $SO(10)$ from the sectors
$b_j$ $(j=1,2,3)$. The $10+1$ and the $E_6$ singlets from the sectors
$b_j+\xi_1$ are replaced by vectorial $16$ of the hidden $SO(16)$ gauge group
from the sectors $b_j+2\gamma$.

The realistic free fermionic models are obtained by moding out the
symmetry with three additional boundary condition vectors that correspond to
Wilson lines in the orbifold formulation. The number of generations is
reduced to three generations, one from each twisted sector $b_1$,
$b_2$ and $b_3$ by reducing the combinatorial factor in Eq. (3) from
eight to one. Each additional vector acts simultaneously on each complex
plane as a $Z_2$ twist, thus reducing the number of generations to exactly
one generation from each sector $b_1$, $b_2$ and $b_3$. Each chiral
generation is obtained from a distinct twisted sector of the orbifold
model and none from the untwisted sector. the reduction to
three generations is correlated with the breaking of the $SO(4)^3$
horizontal symmetries to factors of $U(1)'s$ [\naturalness]. This is however
possible due to the fact that we started from a $Z_2\times Z_2$ orbifold
with the specific choice of radii and background fields, thus producing
the degeneracy of zero modes as in Eq. (3), or alternatively, producing
exactly sixteen fixed points, or eight generations, in each twisted
sector.

The underlying $Z_2\times Z_2$ orbifold compactification has an
important implication for quark and lepton flavor mixing. After
applying the ``Wilson line'' projections each sector $b_1$, $b_2$ and
$b_3$ produces one generation. The $SO(10)$ symmetry is broken to a
subgroup that contains the standard model gauge group [\REVAMP,\ALR] or is
exactly the standard model gauge group times a $U(1)$ [\FNY,\SLM].
The light Higgs doublets are obtained from the Neveu--Schwarz sector and
from a combination of the additional ``Wilson line'', and transform as the
vector representation of $SO(10)$. The standard model gauge group
and its matter content have the traditional $SO(10)$ embedding.
Thus, the weak hypercharge is well defined and unambiguous.
The fermion mass terms in
the low energy effective superpotential are obtained from
renormalizable and nonrenormalizable terms that are invariant under all the
symmetries of the string models [\NRT]. The nonrenormalizable terms become
effective renormalizable terms after giving non--vanishing VEVs to some scalar
singlets in the massless spectrum of the string models. The effective
renormalizable terms are suppressed relative to the terms that are obtained
directly at the cubic level. In this manner one obtains hierarchical
fermion mass and mixing terms [\FM]. The sector $b_3$
produces the lightest generation states while one of $b_1$ and $b_2$
produces the heavy generation states and the other produces the
second generation states [\NRT].
The nonrenormalizable terms that mix between the
generations have a generic form
$$f_if_jhV_i{\bar V}_j{{\phi^n}\over{M^{n+2}}},$$ where $f_i$ and $f_j$ are
fermion states from the sectors $b_i$, $b_j$ with $i\ne j$, $h$ represent the
two light Higgs representations, $V_i$ and ${\bar V}_j$
are two scalars from the
sectors $b_i+2\gamma$ and $b_j+2\gamma$, $\phi^n$ is a combination of scalar
$SO(10)\times SO(16)$ singlets and $M\sim10^{18}GeV$ [\FM,\CKM].
If the states from the sectors $b_j+2\gamma$
get non--vanishing VEVs of order $O({1\over{10}}M)$ semi--realistic
quark mixing matrices can be obtained in these models [\CKM].
We observe that the generic texture of these terms is a result of the
underlying $Z_2\times Z_2$ orbifold compactification.
Namely, the texture of
the mixing terms is of the generic form ${16}_i{16}_j{10}{16}_i{16}_j\phi^n$,
where the first two 16 are in the spinorial representation
of the observable $SO(10)$, the 10 is in the vector representation
of the observable $SO10)$, the last two 16 are in the vector representation of
the hidden $SO(16)$ and $\phi^n$ is a combination of $SO(10)\times SO(16)$
scalar singlets.

In this paper I discussed the orbifold models that correspond to the
realistic models in the free fermionic formulation.
I illustrated in a specific example how the $Z_2\times Z_2$ orbifold model
with a specific choice of background fields and compactification radii
reproduces the spectrum and symmetries of the free fermionic model. I
suggest that the structure of the $Z_2\times Z_2$ orbifold
compactification with standard embedding and at the specific point
in moduli space are the origin of the realistic features of free fermionic
models. In particular, the ``naturalness'' of three generations,
advocated in ref. [\naturalness] is seen to be a result of the
$Z_2\times Z_2$ orbifold compactification with standard embedding and at
the point in compactification space that correspond to the free fermionic
formulation. The free fermionic formulation
correspond to toroidal compactification at the most symmetric point in
compactification space. The $Z_2\times Z_2$ orbifold is the most symmetric
orbifold that one can construct at this point which is consistent with $N=1$
space--time supersymmetry. We are intrigued by the fact that the most
realistic string models to date are constructed at the most symmetric
point in compactification space. Could this be an accident?
A better understanding of the correspondence between the realistic
free fermionic models and other string formulation will
hopefully provide further insight into the realistic features of free
fermionic models.

\bigskip
\centerline{\bf Acknowledgments}
I would like to thank Lance Dixon for useful discussions.
This work is supported by an SSC fellowship.

\refout

\vfill
\eject

\end

This set is referred to as the NAHE{\footnote*{This set was first
constructed by Nanopoulos, Antoniadis, Hagelin and Ellis  (NAHE)
in the construction
of  the flipped $SU(5)$ [\REVAMP].  {\it nahe}=pretty, in
Hebrew.}} set.
The NAHE set is common to all  the realistic models constructed  in the
free fermionic formulation [\REVAMP,\FNY,\ALR,\EU,\TOP,\naturalness]
and is a basic set common to all the models which I discuss.
The sector {\bf S} generates  $N=4$ space--time supersymmetry, which is broken
to $N=2$ and $N=1$ space--time supersymmetry by $b_1$ and $b_2$, respectively.
Restricting $b_j\cdot S=0$mod$2$, and $c\left(\matrix{S\cr
                                                b_j\cr}\right)=\delta_{b_j}$,
for all basis vector $b_j\epsilon{B}$ guarantees the existence of
$N=1$ space--time supersymmetry. The superpartners from a given sector
$\alpha\epsilon\Xi$ are obtained from the sector $S+\alpha$.
The gauge group after the NAHE set is $SO(10)\times
E_8\times SO(6)^3$ with $N=1$ space--time supersymmetry.
The three $SO(6)$
symmetries are horizontal, generational dependent, symmetries.

Models based on the NAHE set correspond to models based on
$Z_2\times Z_2$ orbifold.
This correspondence
is illustrated by extending the  $SO(10)$
symmetry to $E_6$. Adding the vector
$$X=(0,\cdots,0\vert{\underbrace{1,\cdots,1}_{{{\bar\psi}^{1,\cdots,5}},
{{\bar\eta}^{1,2,3}}}},0,\cdots,0)\eqno(3)$$
to the NAHE set, extends the gauge symmetry to
$E_6\times U(1)^2\times SO(4)^3$.
The set $\{{\bf 1},{\bf S},I={\bf 1}+b_1+b_2+b_3,X\}$ produces a
$E_8\times E_8$ toroidal compactification on a $SO(12)$ lattice.
The sectors $b_1$ and $b_2$ correspond to the $Z_2\times{Z_2}$ twist and break
the symmetry to $SO(4)^3\times{E_6}\times{U(1)^2}\times{E_8}$.
The fermionic states $\{\chi_{12},\chi_{34},\chi_{56}\}$ and
$\{{\bar\eta}^1,{\bar\eta}^2,{\bar\eta}^3\}$ give the usual ``standard--
embedding", with
$b(\chi_{12},\chi_{34},\chi_{56})=b({\bar\eta}^1,{\bar\eta}^2,{\bar\eta}^3)$.
The $U(1)$ current of the left--moving $N=2$ world--sheet supersymmetry
is given by $J(z)=i\partial_z(\chi^{12}+\chi^{34}+\chi^{56})$,
and the $U(1)$ charges in the decomposition of $E_6$ under
$SO(10)\times U(1)$ are given by the world--sheet current
${\bar\eta}^1{\bar\eta}^{1^*}+{\bar\eta}^2{\bar\eta}^{2^*}
+{\bar\eta}^3{\bar\eta}^{3^*}$.
The sectors $(b_1;b_1+X)$, $(b_2;b_2+X)$ and $(b_3;b_3+X)$
each give eight $27$ of $E_6$, and correspond to the twisted sectors
of the orbifold model. The $(NS;NS+X)$ sector gives in
addition to the vector bosons and spin two states, three copies of
scalar representations in $27+{\bar {27}}$ of $E_6$. This sector corresponds
to the untwisted sector of the orbifold model.

On the other hand it is general believed that it is, in principle, possible
to construct all string models in the orbifold language.
The number of fixed points in the $Z_2\times{Z_2}$ orbifold on a $SO(12)$
lattice is 48 and matches twice the number of generations in the fermionic
model. The correspondence between the fermionic models and the
$Z_2\times{Z_2}$ orbifold will be discussed further in Ref. [FOE].
The important point to realize is that in the fermionic formulation
the 12 internal fermionic states, $\{y,w\vert{\bar y},
{\bar\omega}\}$, play the role of the six dimensional ``compactified space''
of the orbifold. The boundary conditions, assigned to these internal fermions,
determine many of the properties of the low energy spectrum.

One is drawn to speculate that the realistic
nature of the free fermionic models is not an accident.